# Ultrafast, polarized, single-photon emission from m-plane InGaN Quantum Dots on GaN nanowires

*Tim J. Puchtler\*†, Tong Wang†, Christopher X. Ren‡, Fengzai Tang‡, Rachel A. Oliver‡, Robert A. Taylor†, Tongtong Zhu\*‡*

† Department of Physics, University of Oxford, Parks Road, Oxford, OX1 3PU, UK. ‡ Dept. Materials Science and Metallurgy, University of Cambridge, 27 Charles Babbage Rd., Cambridge, CB3 0FS, UK.

\* tim.puchtler@physics.ox.ac.uk

\* tz234@cam.ac.uk

**Abstract:**

We demonstrate single photon emission from self-assembled m-plane InGaN quantum dots (QDs) embedded on the side-walls of GaN nanowires. A combination of electron microscopy, cathodoluminescence, time-resolved µPL and photon autocorrelation experiments give a thorough evaluation of the QDs structural and optical properties. The QD exhibits anti-bunched emission up to 100 K, with a measured autocorrelation function of $g^{(2)}(0)$ = 0.28 (0.03) at 5 K. Studies on a statistically significant number of QDs show that these m-plane QDs exhibit very fast radiative lifetimes (260 ± 55 ps) suggesting smaller internal fields than any of the previously reported c-plane and a-plane QDs. Moreover, the observed single photons are almost completely linearly polarized aligned perpendicular to the crystallographic c-axis with a degree of linear polarization of 0.84 ± 0.12. Such InGaN QDs incorporated in a nanowire system meet many of the requirements for implementation into quantum information systems and could potentially open the door to wholly new device concepts.

The ability to generate on-demand single-photons is of vital importance to quantum information technology such as quantum cryptography, linear optical quantum computing and quantum metrology.[1–3] Due to their high stability, good repetition rates, and practicable incorporation into cavities and electronically pumped structures,[4] quantum dots (QDs) are ideal candidates for the generation of, and interaction with, single photons. Notably nitride QDs offer the advantages of large exciton binding energies and high band offsets which allow higher temperature operation.[5,6] InGaN QDs in particular allow access to the blue-spectral range ideally suited for free-space communications and efficient fast single-photon detectors. However due to their wurtzite crystal structure the orientation of quantum well (QW) and QD nanostructures relative to the crystal axis is of great importance; most work in the nitrides features growth on the polar (0001) c-plane, causing large in-built fields across the heterostructures leading to decreased oscillator strengths of exciton transitions via the quantum confined Stark effect (QCSE). This limitation on single-photon repetition rates, as well as the unpolarized emission from c-plane QDs, constitute major limitations to their application to quantum information systems. Additionally the exciton binding energy in c-plane structures is theoretically lower than that expected in non-polar orientations.[7] As such there are significant benefits to be realized if QDs can be fabricated on non-polar planes.

Recent advances in the growth of InGaN QDs on $(11\bar{2}0)$ a-plane GaN have demonstrated some of the advantages associated with the reduced internal fields of non-polar planes.[8,9] Additionally the in-plane anisotropy of growth on non-polar planes causes splitting of the upper valence bands and gives these QDs a large degree of inherent polarization,[10] making them candidates for on-chip solutions in polarization encoded quantum information protocols. However fabrication of devices on $(1\bar{1}00)$ m-plane GaN is challenging due to the lack of well-suited lattice-matched substrates.[11] Whilst efficient m-plane QW devices have proven difficult to achieve, requiring expensive free-standing GaN substrates to avoid high defect densities, they have shown better emission properties than even a-plane QWs, having faster radiative recombination times, higher optical gain and an increased degree of linear polarization,[12–15] although the exact reasons for this are not fully understood. However, there has been minimal success in the growth of QDs on the m-plane perhaps because the indium incorporation efficiency is much lower in the m-plane than the c-plane and as such traditional QD growth methods have not translated well onto m-plane substrates.[16] Whilst claims have been made for fabrication of m-plane QDs,[17–19] only the formation of nanostructures on m-plane facets has been demonstrated; there is currently no single-QD spectroscopy which shows optical emission from individual nanostructures, provides evidence of their QD-like behavior, or demonstrates their superior properties relative to c-plane QDs.

Here, we present a novel self-assembled growth method for an m-plane QD system with corresponding structural and optical characterization, and the first direct proof of the quantum nature of emission in such a system. By the use of NWs we allow the relaxation of in-plane strain leading to a reduction in defects, an increase in optical extraction efficiencies from the ends of the NWs. In the future, coupling of the QDs into the photonic mode of the NW is also a possibility. The structure consists of a self-catalytic GaN NW core grown by metal-organic vapor phase epitaxy (MOVPE). We intentionally supply a substantial $SiH_4$ flux during the GaN NW growth step to i) greatly enhance the vertical growth rate, and ii) act as an anti-surfactant that will roughen the NW sidewalls and lead to the formation of three dimensional nanostructures during the growth of 5 InGaN/GaN QW layers which are deposited in a core-shell configuration achieving QD formation. Details of the sample growth are given in the experimental methods section. Formation of m-plane QDs on the sidewalls of hexagonal NWs occurs with both the NW and QD growth performed in a single uninterrupted MOVPE run without the requirement of a previously defined mask, thereby reducing complexity and fabrication time. Such m-plane QD structures can efficiently generate ultrafast linearly polarized single photons up to 100 K which can potentially be coupled into the photonic modes of the hexagonal nanowires.

Structural properties of the NWs have been assessed using scanning electron microscopy (SEM) with an attached cathodoluminescence detector (SEM-CL) allowing emission spectra to be collected with high spatial resolution from regions excited locally by the electron beam. Variation in emission along the NW can be seen in Figure 1 (a,b). We see that the InGaN emission at the base and middle of the NW is consists of many sharp peaks attributed to separate localization centers. Emission becomes stronger and more uniform towards the NW tip, displaying QW-like emission centered at 400 nm, having minimal spectral overlap with the sharp emission features observed.

To gain further insight into the InGaN layer morphology, high-resolution transmission electron microscopy (TEM) imaging has been performed on cross-sectional lamella prepared by focused ion beam (FIB) milling. It reveals that the distribution of indium is of a uniform thickness towards the top of the NWs (Figure 1c), consistent with the SEM-CL spectra showing QW-like emission. However, throughout most of the NW the InGaN layers appear non-uniform, with variations in both thickness and concentration. Energy-dispersive X-ray (EDX) spectroscopy performed concurrently shows the presence of high indium-content structures (approximately 5 nm high x 15 nm across in size) throughout these non-uniform layers (Figure 1d,f) which, when combined with the SEM-CL spectra showing isolated sharp peaks, confirms the formation of QDs.

A silicon flux during growth has been shown to greatly enhance the vertical growth rate of GaN NWs. Tessarek *et al.* suggested that during this process a $SiN_x$ layer or highly Si-doped GaN surface layer can be incorporated and formed on the m-plane sidewalls of the NWs due to 1) the presence of Ga-droplets on top of the GaN NWs during growth which attract the Ga atoms and 2) the low solubility of Si and N in liquid Ga.[20] In the growth of our NWs, we supplied a constant $SiH_4$ flux (0.2 µmol/min) during the GaN NW growth in order to increase the vertical growth rate. This $SiH_4$ was stopped prior to the InGaN QW growth, to encourage the formation of a core shell structure. The carrier gas was then changed to $N_2$ and the growth temperature was ramped down to the QW growth temperature with a constant supply of ammonia. We suggest that this process leads to a variation of $SiN_x$ coverage along the nanowire length, with more $SiN_x$ at the bottom of the nanowire which had been exposed to $SiH_4$ for a longer time. The formation of a $SiN_x$ layer on planar surfaces is known to lead to a transition from 2D to 3D growth.[21] This so-called 'anti-surfactant effect' of Si on NW's surface is postulated to disturb the growth of the GaN beneath the first InGaN QW layer (perhaps by the formation of 3D islands which then partially coalesce). This results in a rougher surface towards the bottom of the NW for the subsequent InGaN/GaN QWs to grow on. Higher InGaN growth rate (thickness), and higher local indium incorporation (composition) due to the presence of nanoscale facets as confirmed by the TEM and EDX analysis (Figure 1c and 1d),[22] lead to the formation of m-plane QD-like structures on the NW's sidewalls. Additional structural information is presented in the Supporting Information (SI). And a more detailed structural and chemical analysis of the effect of $SiN_x$ on InGaN/GaN core-shell NW growth will be presented elsewhere.

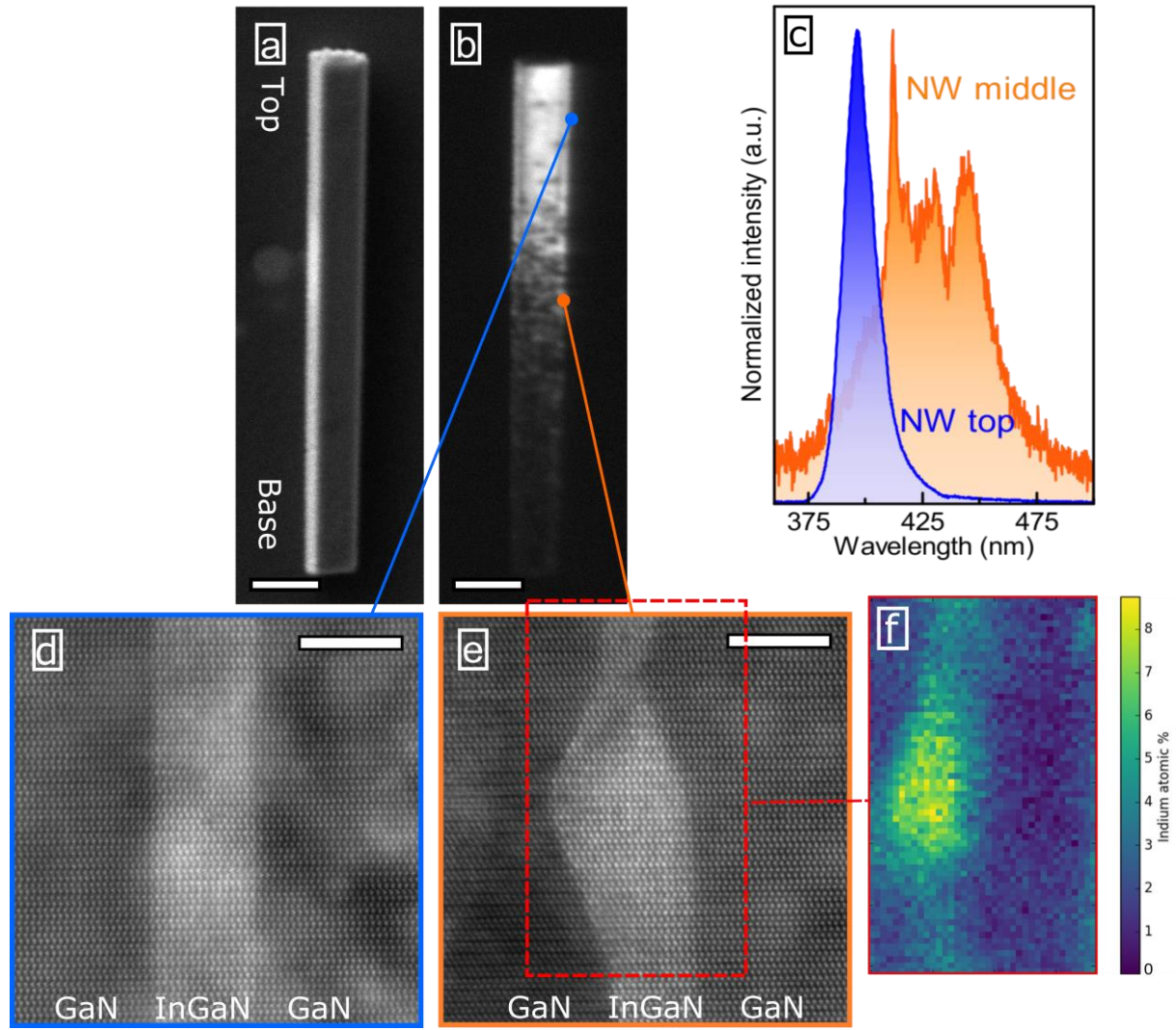

**Figure 1.** (a) SEM image of a NW with (b) corresponding CL image showing a transition from fractured to uniform InGaN emission towards the top of the NW. Scale bar represents 1 μm. (c) Representative CL emission spectra for these two regions, showing QW-like emission towards the top of the NW and sharper peaks within the fractured region. Corresponding TEM images of the InGaN layers (d) at the top and (e) from the middle of the NW showing QW- and QD-like morphologies respectively. Scale bar represents 3 nm. Indium clustering is easily identified in (f) corresponding EDX spectra.

Optical emission characteristics of the NWs were measured using micro-photoluminescence (µPL) under two-photon excitation (λ = 800 nm), as this has been shown to give a relatively larger absorption cross-section for objects of increased quantum confinement and hence increases QD signal relative to the non-uniform QW background.[23] Consistent, smooth QW-like emission typically centered at 415 ± 2 nm is observed from the top of the NW, whilst sharp emission peaks are observed at other positions along the NW (Figure 1e), showing strong similarities to the emission observed under SEM-CL. Many of these peaks are present in each NW with a relatively wide distribution of emission wavelengths (average 473 nm, standard deviation $\sigma$ = 28 nm, for 66 peaks examined).

Power dependent measurements have been performed to allow identification of any biexcitons present. The power dependence of the QD emission varies approximately with the square of the excitation power ($\propto P^{2.1}$, see Figure 2a inset) as expected for an exciton created via a two-photon absorption process, with only two biexcitons observed having power dependence to the power 4.4. The rarity of observed biexcitons may suggest a large biexciton binding energy, or simply a weak peak which is hidden by the background emission at higher powers beyond the saturation of the exciton peak (6 mW in Figure 2). The emission wavelength appears relatively insensitive to excitation power with no clear correlation observable for studied QDs. Whilst some degree of spectral diffusion is observed (typically peak wavelength varies by ~ 0.05 nm over a period of 10s of seconds), which may be concealing a power dependent emission shift, the small magnitude strongly suggests that the in-built fields which are screened at higher excitation power densities for c-plane QDs are minimal in the QDs currently observed.

In order to prove that the emission features come from individual quantum emitters, autocorrelation measurements have been performed using a Hanbury Brown and Twiss setup.

QD emission is spectrally filtered using a pair of tunable band-pass filters, although some degree of background emission still remains (illustrated in Figure 2a). Whilst the background emission cannot be completely removed by spectral filtering, clear antibunching is observed in the second-order autocorrelation function (Figure 2b) giving a correlation signal at zero time delay of $g^{(2)}(0) = 0.28$. This value can be corrected by removing the effect of the uncorrelated background emission according to $\left(g^{(2)}_{\text{cor}}(0) - 1\right)\rho^2 = (g^{(2)}_{\text{raw}}(0) - 1)$ where $\rho$ is the ratio of QD signal to total intensity measured.[24] Doing so gives a value of $g^{(2)}_{\text{cor}}(0) = 0.03$. This non-zero value may be explained by re-excitation of the QD from background emission during a single excitation pulse, but is more likely caused by inaccurate estimations of $\rho$: difficulties arise as the tunable filter's transmission is not a perfect boxcar function, having a small but significant transmission beyond the intended transmission window. Errors may also arise in the corrected value if the QD peak shape is incorrectly fitted, such as the influence of broadening mechanisms or phonon-sidebands, which may become significant at higher temperatures. (See SI for information on estimations of $\rho$). This is evident in additional autocorrelation measurements performed at 100 K which give values of $g^{(2)}_{\text{raw}}(0)$ of 0.49 and $g^{(2)}_{\text{cor}}(0)$ of 0.18 respectively (Figure 2c). Whilst the value of $\rho$ becomes lower and more difficult to estimate accurately as the QD thermally broadens, antibunching behavior remains at this significantly elevated temperature. We should also note that the sample was grown with 5 InGaN layers in order to increase the likelihood of QD formation. However, given that most nanowires contained several measurable QDs, a reduction in background emission can be achieved by simply reducing the number of layers present.

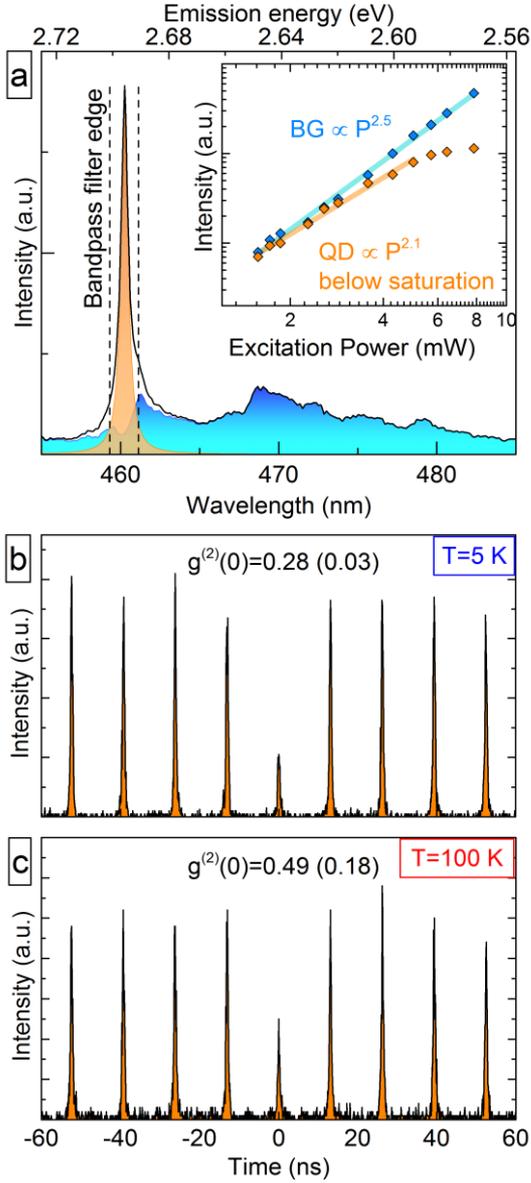

**Figure 2.** (a) Example emission spectrum from a QD at 5 K showing a single peak attributed to a QD and a wide background emission attributed to fractured QW in its vicinity. Example fits of the QD, background and filters used for background correction in autocorrelation measurements are shown. (Inset) Emission intensity follows a near-quadratic power dependence until saturation, as expected for a two-photon absorption process. Autocorrelation measurements of a QD at (a) 5 K showing significant antibunching of $g^{(2)}(0) = 0.27$ and at (c) 100 K showing antibunching behavior with $g^{(2)}(0) = 0.49$.

Whilst the formation of QDs has been seen in TEM and sharp peaks have been observed in µPL, one might consider the possibility that the two are not directly correlated; it is possible that the emission peaks related to excitons bound to defects or impurities present in the NWs, an especially important consideration given the role of high silicon levels plays in the formation mechanism of the QDs themselves. However, firstly we should note that the excitation power near saturation does not vary with the square-root of excitation density and as such suggests that the emission is not defect-related.[25] In order to confirm this assessment, temperature dependent µPL measurements have been performed on a representative emission peak, as the emission peaks from QDs and defects behave differently with temperature. Example spectra for a single emission peak from 4 – 220 K may be seen in Figure 3 with the emission wavelength increasing smoothly, with a rate increasing with temperature as expected for semiconductor band-gaps following the Varshni equation.[26] Alternatively, for defect related emission, we would expect to observe a progression in peak intensity representing the transition from bound- to free- excitons as we increasingly thermalize the donor-bound excitons.[27] Indeed the wavelength change fits almost perfectly ($R^2 = 0.996$) to the O'Donnell-Chen model of temperature dependence of band gaps (see SI for details), and as such we can conclude that the emission features presented do correspond to semiconductor localization centers such as those observed in TEM.

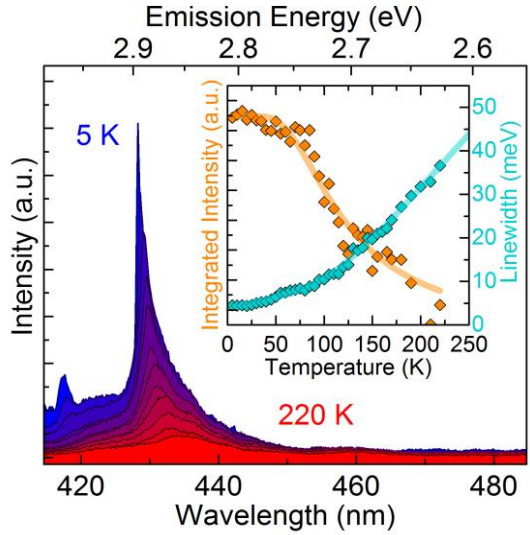

**Figure 3.** Variation in QD emission with temperature up to 220 K, showing a gradual red-shift and (inset) decrease in intensity and increase in linewidth with increasing temperature as expected for QD emission rather than defect-related emission.

The variation of linewidth with increasing temperature gives us insight into the influence of phonons on these m-plane QDs. We expect significant acoustic-phonon broadening in the nitrides,[28] whilst having minimal coupling to optical phonons at the range of temperatures measured due to the large optical phonon energy in GaN (92meV).[29] The linewidth $\Gamma$ of QDs fit well with the standard equation for bulk or QW semiconductors,[30] $\Gamma(T) = \Gamma_0 + \gamma_a T + \gamma_b \exp(-E_a/k_B T)$ with terms representing the zero-phonon linewidth (ZP-linewidth), acoustic phonon coupling and delocalization of carriers from the QD respectively. We can see in Figure 3 (inset) that the acoustic phonon coupling is indeed the most significant for temperatures up to ~ 100 K, giving a coupling constant of 50 ± 7 µeV and $\Gamma_0$ of 3.7 ± 0.4 meV. These linewidths are several times higher than many other reported values for InGaN QDs.[28,30,31] We have attributed the larger-than-expected ZP-linewidth to the likely significant presence of spectral diffusion in the sample, given the probable high density of point defects associated with the Si dopants, which act as charge trapping sites in the regions in which QDs are formed, as discussed previously. It would seem consistent that we observe a correspondingly larger acoustic phonon coupling constant, as the magnitude of the spectral diffusion is also expected to increase with temperature.[32] Therefore this value of $\gamma_a$ is not purely the acoustic coupling, but contains a measure of the increased thermal movement of carriers as well.

The reduction in integrated intensity from the QD has been plotted in Figure 3 (inset) and fits well with an Arrhenius-type equation based on a single carrier escape energy, $I = I_0/(1 + A_1 \exp(-E_1/k_B T))$. The energy $E_1$ representing the energy required for an exciton to escape the QD localization is extracted as $35 \pm 1$ meV. This corresponds excellently to the extracted value of the localization depth used in fitting the linewidth previously which yielded $E_a = 38 \pm 2$ meV, with these values being slightly larger than those previously reported in literature for c-plane InGaN QDs, and hence explaining the higher operational temperature we have observed compared to those studies.[30,33]

Note that due to the fast lifetimes the homogeneous linewidth of the QDs is beyond our resolution limit. Hence to extract FWHM and amplitude data for Figure 3 we have used Gaussian functions to fit our emission peaks, representing inhomogeneous broadening mechanisms such as spectral diffusion.

To gain insights into the exciton recombination dynamics of the QDs, time resolved measurements have been performed on 32 spectrally isolated QDs. Excitation power was chosen to be well below saturation for each QD to reduce any effect of carrier-generated internal fields. Decay traces fit well with two-component exponential decays, with one strong fast component attributed to the QD and a significantly weaker (by orders of magnitude) slower component. As such, and noting the QD is weakly pumped whilst the emission is spatially and spectrally isolated, the stretching parameter often used to account for temporally-varying lifetimes or ensembles of emitters (the Kohlrausch function) is not required.[34] To allow measurement of very fast lifetimes, fitting of the bi-exponential decay is convolved with the detector's Gaussian instrument response function (IRF, ~150 ps width). An example may be seen in Figure 4a.

The lifetime of the QD exciton was found to be as low as 170 ps, with an average of 260 ± 55 ps for the 32 QDs measured. This value is much lower than typical lifetimes for c-plane QDs (typically 1 – 10 ns)[35,36] supporting the assertion that the non-polar orientation of these QDs reduces the effect of their built-in fields, reducing the QCSE and increasing oscillator strength. Interestingly these lifetimes are also significantly faster than those reported in InGaN QDs grown on the a-plane, typically of ~ 500 ps,[8,9] suggesting that these QDs better avoid residual in-built fields which may be caused by any semi-polar facets than their a-plane counterparts. Indeed, these lifetimes are less than those observed in quantum disks which have claimed to be free from internal fields (~300 ps)[37].

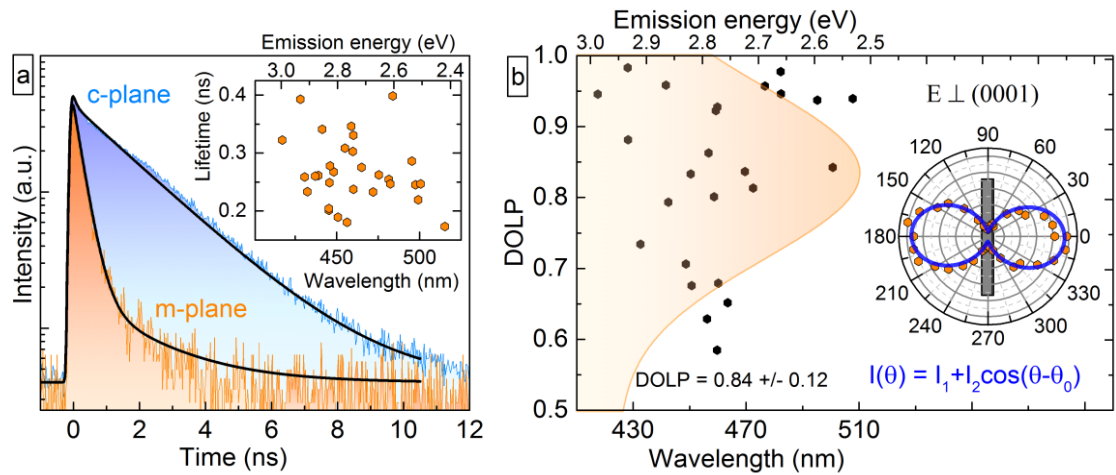

**Figure 4.** (a) Time-resolved PL decay traces of the m-plane QDs studied compared to the longer lifetime peaks sometimes present when exciting the tip of the NW attributed to c-plane QDs. Average lifetimes are 260 ± 55 ps, suggesting minimal internal fields within the QDs which show (inset) no correlation to emission wavelength. (b) Degree of linear polarization

measurements against emission wavelength, showing a high average DOLP of 0.84 with no wavelength dependence. The shaded Gaussian is a guide for the eye representing the distribution of DOLPs measured. (Inset) An example 360° polarization plot showing the orientation of polarization relative to the NW, with all QDs measured exhibiting polarization orthogonal to the NW growth direction.

The lack of significant fields within the QDs is further supported, albeit tentatively, by the lack of wavelength dependence of the lifetime of the excitons; typically exciton binding energy decreases with increasing QD size, decreasing emission energy.[38] Similarly we expect an increase in emission lifetime as the electron-hole wavefunction overlap is reduced, with a magnitude dependent on the internal field and geometry of the QD. The resulting inverse correlation of emission energy to lifetime strongly depends on the presence of built-in fields,[34] assuming isolated transitions are measured. Whilst our QD growth relies on a stochastic formation process which results in a range of QD sizes and compositions, we see no such relationship for our QDs (Figure 4, inset, gives an $R^2$ of 0.02 to a linear fit).

We should note that for several NWs, QD emission may be observed with much longer lifetimes (~2 ns) exclusively when excited at their top end. We believe these may indicate the presence of c-plane QDs formed at the apex of the hexagonal pyramidal facets as previously reported for InGaN QWs grown atop pyramidal structures.[39]

Finally, we investigate the polarization properties of the QDs. In m-plane heterostructures we expect a high degree of in-plane polarization, even without the presence of anisotropic in-plane strain or in-built fields, as the crystallographic asymmetry lifts the degeneracy of the top two p-like valence band states.[40,41] A high degree of polarization has already been repeatedly demonstrated in m-plane QW structures.[42,43] However, it is possible that shape anisotropy of these QDs could randomize the polarization via the non-equal strain fields affecting the A and B valence band states differently.[44,45]

An example of the QD intensity as a function of polarization angle is presented in Figure 4b (inset), fitted with $I(\theta) = I_1 + I_2 \cos(\theta - \theta_0)$ with $\theta_0$ as the angle of polarization, $I_1$ and $I_2$ as the minimum and maximum intensities respectively. Polarization measurements performed on 26 QDs are presented in Figure 4 and show a high degree of linear polarization, defined as DOLP $= (I_2 - I_1)/(I_1 + I_2)$, measured at 0.84 ± 0.12. For all QDs measured the angle of polarization is aligned orthogonal to the NWs, hence in the E ⊥ c direction as expected from m-plane QWs. We also note that the DOLP has no dependence on the wavelength of QD emission (a linear fit gives an $R^2$ of 0.03). We can therefore conclude that the polarization is robust to any shape anisotropy present in these QDs.

In conclusion we have presented the first successful growth of m-plane QDs using a novel method in which the QDs self-assemble owing to higher InGaN growth rate and indium content on the nanofacets induced by $SiN_x$ on the sidewalls of hexagonal GaN NWs, and have demonstration of their anti-bunched emission. We have performed detailed analysis on the optical properties of these QDs, and we conclude that they show very fast lifetimes (as low as 170 ps) suggesting the presence of only minimal internal fields. The QDs also show a high degree of linear polarization (DOLP ~ 0.84) and antibunching $g^{(2)}(0)$=0.28 (0.03), although the QDs also exhibit a high level of spectral diffusion likely caused by higher impurities in the material on which the QDs form.

**Experimental Methods:**

Sample Growth:

The NWs were grown by metal-organic vapor phase epitaxy in a 6 x 2 in. Thomas Swan close-coupled showerhead reactor on 2" c-plane sapphire substrates using trimethylgallium, trimethylindium, and ammonia as precursors, hydrogen as carrier gas and $SiH_4$ as source of silicon. After annealing the sapphire substrate at 1050 °C, a low temperature nucleation GaN layer (~5 nm) was deposited at 550 °C and annealed in $H_2$ at 1000 °C to act as a mask for the subsequent NW growth. The NWs were then grown at 1000 °C with a reactor pressure of 300 Torr and a V/III ratio of 50, during which a constant $SiH_4$ flow of (0.2 μmol/min) was supplied to enhance the vertical growth rate. A five-period core-shell InGaN/GaN multiple quantum wells were grown on the GaN core NWs using a V/III ratio of 5000 and $N_2$ as the carrier gas, whereby the InGaN was grown at 730 °C and the GaN barriers were grown at 850 °C. Samples have been harvested and redeposited on a silicon substrate for analysis.

Characterization:

Cathodoluminescence measurements were performed at 30 K using a Philips XL30 SEM operating at 3 kV, which equipped with a Gatan MonoCL4. Site-specific TEM (Transmission Electron Microscopy) samples were prepared from NWs using the dual beam FIB-SEM (Focused ion beam – Scanning Electron Microscopy) based lift-out technique (FEI Helios NanoLab$^{TM}$). Atomic resolution HAADF (High-Angle Annular dark field)-STEM images were taken using a spherical aberration-corrected TEM (FEI Titan$^3$ 80-300) at 300 kV with screen currents between 50-250 pA, whereas chemical composition analysis was performed by an analytical TEM (FEI Tecnai Osiris) fitted with four EDX spectrometers operating at 200 kV. TEM analysis was carried out with electron beam parallel to the ⟨11$\bar{2}$0⟩ direction. μPL measurements were performed using a mode-locked Ti:Sapphire laser emitting at 800 nm (1.5 ps pulse length, 76 MHz repetition rate) through a 0.7 NA objective (spot size ~ 0.6 μm$^2$) with the emission passing through the same objective before being passed to a 0.5m spectrometer with a 1200 lpmm grating (maximal resolution of 0.04 nm). The sample is held in a closed-cycle cryostat (AttoDRY800). Lifetime measurements were performed passing a wavelength selected output from the spectrometer to a photomultiplier tube (PMT, instrument response function width ~ 150 ps) connected to a time-correlated single-photon counting module (binning resolution 25 ps) triggered by the Ti:Sapphire pulses. Auto-correlation measurements were performed passing the QD signal through a pair of tunable bandpass filters (Semrock Versachrome) to a pair of PMTs through a 50:50 beamsplitter in a Hanbury Brown and Twiss configuration. Polarization measurements were performed using a half-wave plate and linear polarizer in front of the spectrometer.

**Author Contributions**

The experiments were conceived by T.J.P and T.Z., optical measurements performed by T.J.P. and T.W., SEM-CL measurements performed by T.Z., samples grown by T.Z., TEM sample preparation and imaging by C.X.R. and F.T., data analysis performed and paper written by T.J.P., growth and structural analysis supervised by R.A.O. and optical characterization supervised by R.A.T.

**Funding Sources**

This research was supported by the Engineering and Physical Sciences Research Council (EPSRC) UK (Grant No. EP/M012379/1 and EP/M011682/1). The microscopy studies were supported by the European Research Council under the European Community's Seventh Framework Programme (FP7/2007-2013)/ERC grant agreement no 279361 (MACONS).